\documentclass[12pt,preprint,preprintnumbers,showpacs,nofootinbib]{revtex4}

\usepackage[dvips]{graphicx}

\input{epsf.sty}

\usepackage{graphicx}
\usepackage{epsfig}
\usepackage{dcolumn}
\usepackage{bm}
\usepackage{amssymb}

\renewcommand{\thefootnote}{\fnsymbol{footnote}}

\def\hide#1{[hidden stuff]}

\def\beq{\begin{equation}}
\def\eeq{\end{equation}}
\def\eq{\end{equation}}
\def\to{\rightarrow}

\def\mEt{\mbox{${\hbox{$E$\kern-0.6em\lower-.1ex\hbox{/}}}_T$}\, } 

\def\bsg{\ifmmode B_d\to X_s\gamma\else $B_d\to X_s\gamma$\fi}
\def\bsglue{\ifmmode B_d\to X_s\, g\else $B_d\to X_s\, g$\fi}
\def\bsll{\ifmmode B_d\to X_s\ell^+\ell^-\else $B_d\to X_s\ell^+\ell^-$\fi}
\def\bstt{\ifmmode B\to X_s\tau^+\tau^-\else $B\to X_s\tau^+\tau^-$\fi}
\def\shat{\ifmmode \hat{s}\else $\hat{s}$\fi}

\def\bphik{\ifmmode B_d\to \phi K_s\else $B_d\to \phi K_s$\fi}
\def\bbarphik{\ifmmode \bar{B_d}\to \phi K_s\else $\bar{B_d}\to \phi K_s$\fi}
\def\bsmix{\ifmmode B_s \bar{B}_s\else $B_s \bar{B}_s$\fi}
\def\bdmix{\ifmmode B_d \bar{B}_d\else $B_d \bar{B}_d$\fi}
\def\bclnu{\ifmmode B_d\to X_c e \bar{\nu}\else $B\to X_c e \bar{\nu}$\fi}
\def\bjpsik{\ifmmode B_d\to J/\psi~K_s\else $B_d\to J/\psi~K_s$\fi}
\def\bsee{\ifmmode B_d\to X_s e^+ e^-\else $B_d\to X_s e^+ e^-$\fi}
\def\bsmumu{\ifmmode B_d\to X_s\mu^+\mu^-\else $B_d\to X_s \mu^+\mu^-$\fi}

\newcommand{\newc}{\newcommand}

\newc{\asusy}{\delta a^{\rm SUSY}_\mu}
\newc{\lcal}{\int {\cal L}dt}
 
\newc{\LSP}{{\chi^0_1}}
\newc{\stauR}{{\tilde \tau_R}}
\newc{\stau}{{\tilde \tau_1}}
\newc{\mstop}{m_{\tilde{t}}}
\newc{\mHpm}{m_{H^\pm}}
\newc{\gsim}{\lower.7ex\hbox{$\;\stackrel{\textstyle>}{\sim}\;$}}
\newc{\lsim}{\lower.7ex\hbox{$\;\stackrel{\textstyle<}{\sim}\;$}}
\newc{\ie}{{\it i.e.}}          
\newc{\etal}{{\it et al.}}
\newc{\eg}{{\it e.g.}}          
\newc{\kev}{\hbox{\rm\,keV}}            
\newc{\mev}{\hbox{\rm\,MeV}}            
\newc{\gev}{\hbox{\rm\,GeV}}            
\newc{\tev}{\hbox{\rm\,TeV}}
\newc{\xpb}{\hbox{\rm\, pb}}
\newc{\xfb}{\hbox{\rm\, fb}}

\newc{\mtop}{m_t}
\newc{\mbot}{m_b}
\newc{\mz}{m_Z}
\newc{\mw}{M_W}
\newc{\alphasmz}{\alpha_s(m_Z^2)}
\newc{\swsq}{\sin^2\theta_W}
\newc{\tw}{\tan\theta_W}
\newc{\cw}{\cos\theta_W}
\newc{\sw}{\sin\theta_W}
\newc{\BR}{\hbox{\rm BR}}
\newc{\zbb}{Z\to b\bar}
\newc{\Gb}{\Gamma (Z\to b\bar b)}
\newc{\Gh}{\Gamma (Z\to \hbox{\rm hadrons})}
\newc{\rbsm}{R_b^\hbox{\rm sm}}
\newc{\rbsusy}{R_b^\hbox{\rm susy}}
\newc{\drb}{\delta R_b}

\newc{\sgn}{\mbox{sgn}}
\newc{\tbeta}{\tan\beta}
\newc{\uL}{{\tilde u_L}}
\newc{\uR}{{\tilde u_R}}
\newc{\cL}{{\tilde c_L}}
\newc{\cR}{{\tilde c_R}}
\newc{\tL}{{\tilde t_L}}
\newc{\tR}{{\tilde t_R}}
\newc{\dL}{{\tilde d_L}}
\newc{\dR}{{\tilde d_R}}
\newc{\sL}{{\tilde s_L}}
\newc{\sR}{{\tilde s_R}}
\newc{\bL}{{\tilde b_L}}
\newc{\bR}{{\tilde b_R}}
\newc{\eL}{{\tilde e_L}}
\newc{\eR}{{\tilde e_R}}
\newc{\mhp}{m_{H^\pm}}
\newc{\mhalf}{m_{1/2}}
\newc{\emt}{{e/\mu /\tau}}

\newc{\lR}{\tilde{l}_R}
\newc{\lL}{\tilde{l}_L}
\newc{\nL}{\tilde{\nu}_L}
\newc{\na}{\chi^0_1}
\newc{\nb}{\chi^0_2}
\newc{\nc}{\chi^0_3}
\newc{\nd}{\chi^0_4}
\newc{\ca}{\chi^{\pm}_1}
\newc{\cb}{\chi^{\pm}_2}
\newc{\camp}{\chi^\mp_1}
\newc{\cbmp}{\chi^\mp_1}
\newc{\capos}{\chi^{+}_1}
\newc{\caneg}{\chi^{-}_1}
\newc{\phit}{\phi_t}
\newc{\phib}{\varphi_b}
\newc{\phiew}{\phi_{ew}}
\newc{\htz}{h^0_t}
\newc{\hbz}{h^0_b}
\newc{\hewz}{h^0_{ew}}
\newc{\hsmz}{h^0_{sm}}
\newc{\huz}{h^0_u}
\newc{\hsusyz}{h^0_{susy}}

\newcommand{\drawsquare}[2]{\hbox{%
\rule{#2pt}{#1pt}\hskip-#2pt
\rule{#1pt}{#2pt}\hskip-#1pt
\rule[#1pt]{#1pt}{#2pt}}\rule[#1pt]{#2pt}{#2pt}\hskip-#2pt
\rule{#2pt}{#1pt}}

\newc{\Dal}{\drawsquare{7}{0.6}}

\def\beq{\begin{equation}}
\def\eeq{\end{equation}}
\def\bea{\begin{eqnarray}}
\def\eea{\end{eqnarray}}
\def\mhs{m^2_{h}}

\def\mones{m^{(m)\, 2}}
\def\mtwos{m^{(n)\, 2}}
\def\mone{m^{(m)}}
\def\mtwo{m^{(n)}}
\def\mns{m^2_{\nu}}
\def\mn{m_{\nu}}
\catcode`@=11
\long\def\@caption#1[#2]#3{\par\addcontentsline{\csname
  ext@#1\endcsname}{#1}{\protect\numberline{\csname
  the#1\endcsname}{\ignorespaces #2}}\begingroup
    \small
    \@parboxrestore
    \@makecaption{\csname fnum@#1\endcsname}{\ignorespaces #3}\par
  \endgroup}
\catcode`@=12

\begin{document}

\begin{flushright}
MSUHEP-031219 \\
hep-ph/0312339
\end{flushright}

\title{
Constraints on Large Extra Dimensions with \\
Bulk Neutrinos}

\author{Qing-Hong Cao} \email{cao@pa.msu.edu}
\author{Shrihari Gopalakrishna} \email{shri@pa.msu.edu}
\author{C.--P. Yuan} \email{yuan@pa.msu.edu}

\affiliation{
\vspace*{2mm}
{Department of Physics and Astronomy, \\ 
   Michigan State University, \\
     East Lansing, MI 48824, USA. \\ }}

\vspace{0.15in}
 
\begin{abstract}

We consider right-handed neutrinos propagating in $\delta$ (large) extra dimensions, whose only 
coupling to Standard Model fields is the Yukawa coupling to the left-handed neutrino and the Higgs boson.
These theories are attractive as they can explain the smallness of the neutrino mass, as
has already been shown. We show that if $\delta$ is bigger than two, there are strong constraints 
on the radius of the extra dimensions, resulting from the experimental limit on the probability of
an active state to mix into the large number of 
sterile Kaluza-Klein states of the bulk neutrino. We also calculate the bounds on the radius 
resulting from requiring that perturbative unitarity be valid in the theory, in an imagined 
Higgs-Higgs scattering channel.
\end{abstract}



\maketitle


\setcounter{footnote}{0}
\renewcommand{\thefootnote}{\arabic{footnote}}


\baselineskip=18pt

\section{Introduction}

The Standard Model (SM) of high energy physics suffers from the gauge hierarchy problem, 
which is the fine tuning required to maintain a low electroweak scale ($M_{EW} \sim 10^3$~GeV)
in the presence of another seemingly fundamental scale, the Planck scale (the scale of gravity, 
$M_{pl}\sim 10^{19}$~GeV).
Supersymmetry, technicolor and more recently extra (space) dimensions have been proposed to 
address the hierarchy problem. 

Recent neutrino oscillation experiments have suggested a non-zero neutrino mass, with the best
fit values of the mass differences and mixing angles given 
by~\cite{Fukuda:1998mi,Ahmad:2002ka,Wolfenstein:1977ue,Bahcall:2002hv}
\footnote{In this work, we will not address the LSND result~\cite{Aguilar:2001ty}.}
\bea
\Delta m^2_{\rm Solar} &=& 7\times 10^{-5}~{\rm eV^2} \ , \ \ \ \tan^2\theta_{Solar} = 0.4 \ , \nonumber \\
\Delta m^2_{\rm Atm} &=& 2.5\times 10^{-3}~{\rm eV^2} \ , \ \ \ \tan^2\theta_{Atm} = 1 \ .
\label{DMSQ.EQ}
\eea
The oscillation between the three active flavors,  $\nu_e$,$\nu_\mu$,$\nu_\tau$, in the SM, 
accommodates this satisfactorily, with 
$\Delta m^2_{\rm Solar} = |m_1^{2}-m_2^{2}|$ and $\Delta m^2_{\rm Atm} = |m_2^{2}-m_3^{2}|$, the 
$m_i$ being the physical masses.
If the $m_i$ are also assumed to be of the same order of magnitude as the mass differences, 
it is quite challenging to explain why it is that the neutrinos are so light compared to the other leptons. 

It has been shown~\cite{Arkani-Hamed:1998rs,Antoniadis:1998ig} that if 
there are other Large Extra Dimensions (LED) in addition to our usual four space-time dimensions, 
we could potentially solve the gauge hierarchy problem. It was then pointed 
out~\cite{Dienes:1998sb,Arkani-Hamed:1998vp,Dvali:1999cn} that the smallness of the neutrino 
mass is naturally explained\footnote{We note here that the 
conventional see-saw mechanism to explain the smallness of 
the neutrino mass is equally appealing, but we will not consider it in this work.} 
if right-handed neutrinos that propagate in some $\delta$ number of these extra 
dimensions are introduced. We will refer to such neutrinos, which are SM gauge singlets, as 
``bulk neutrinos'', as is the usual practice. Various aspects of theories with bulk neutrinos have been
analyzed in Ref.~\cite{Mohapatra:1999zd}. 

If bulk right-handed neutrinos are responsible for the smallness of the 
neutrino mass, we ask in this paper, what the constraints on such a model might be.
We will find that the experimental constraints on the probability of an active neutrino to oscillate 
into sterile bulk neutrinos, can give us a lower bound on the inverse radius ($1/R$) of such
extra dimensions, especially if $\delta > 2$. 
We will also require that perturbative unitarity in 
two-to-two scattering of Higgs bosons be preserved in the theory. We will include
the tower of Kaluza-Klein states as intermediate states, and again find that for 
$\delta>2$ it results in a strong bound on $1/R$. 

We will take the view, as in Ref.~\cite{Davoudiasl:2002fq}, that the standard three-active-flavor 
oscillation explains the data in Eq.~(\ref{DMSQ.EQ}), and that 
the mixing to sterile bulk neutrinos are small enough to evade experimental constraints. 
Here we extend the analysis to $\delta>1$ and show that strong bounds on $1/R$ can result
in that case. However, a precise calculation of the bound will not be possible when 
$\delta>2$ due to sensitivity on the cutoff scale, implying a dependence on how one completes 
the extra dimensional (effective) theory that we will work with. As pointed out in 
Ref.~\cite{Davoudiasl:2002fq} an alternative approach~\cite{Dienes:1998sb,Dvali:1999cn}, 
wherein Eq.~(\ref{DMSQ.EQ}) is explained by the oscillation of the active species predominantly into 
sterile bulk neutrinos, appears to be disfavored by the Sudbury Neutrino 
Observatory~(SNO)~\cite{Ahmad:2002ka} neutral current data. 

In order to constrain theories with neutrinos in the bulk, other processes have also 
been considered in the literature~\cite{Ioannisian:1999cw} including muon lifetime, pion decay, 
flavor violation, beta decay in nuclei, muon $g-2$, supernova energy loss and big bang nucleosynthesis. 
Some phenomenological implications are also considered in Ref.~\cite{DeGouvea:2001mz}. 

The rest of the paper is organized as follows. 
We will introduce the extra dimensional theory with bulk neutrinos in Sec.~\ref{BNLED.SEC}, 
write down the equivalent four dimensional Kaluza-Klein theory with particular focus on
the interaction of the right-handed bulk neutrino with the Higgs field and the left-handed
neutrino, and diagonalize the neutrino mass matrix. In Sec.~\ref{OSCON.SEC}, we will find the 
bounds on the radius of the extra dimensions from limits on neutrino oscillation into 
sterile states, and in Sec.~\ref{UNICON.SEC}, from unitarity considerations in Higgs-Higgs scattering. 
Our conclusions are given in Sec.~\ref{CONCL.SEC}. We will present an alternate approach to 
the diagonalization of the neutrino mass matrix in Appendix~\ref{MDDS.APP}, and give in 
Appendix~\ref{IPV.APP} some detailed formulas that we use in deriving the unitarity bound.

\section{Right-handed Neutrinos in Extra Dimensions}
\label{BNLED.SEC}
To address the gauge hierarchy problem, Arkani-Hamed, Dimopoulos and Dvali 
(ADD)~\cite{Arkani-Hamed:1998rs} postulate that the 
Standard Model (SM) fields are confined to a 4 dimensional (4-D) sub-space (brane) in an 
extra dimensional world of $4+n$ total dimensions. 
ADD take the view that the only fundamental scale in nature is $M_*$, which is of the order of 
$M_{EW}$, and the apparent 4-D gravity scale ($M_{pl}$) is then given by
\bea
M_{pl}^2 = M_*^{2+n} V_n,
\label{ADDMPL.EQ}
\eea
where, $V_n$ is the volume of the (compact) extra dimensional space. In the simple case of 
each of the compact extra dimensions being of equal radius $R^\prime$, we have $V_n \sim {R^\prime}^n$. 
Thus ADD argue that $M_{pl}$ appears to be a large scale from a 4 dimensional perspective 
simply because the volume $V_n$ is large. In other words, the explanation of why $M_{pl}$ is 
large is recast to stabilizing $R^\prime$ at a large value, so that $V_n$ is large.

It should be pointed out that for a given $M_*$, if the $n$ compact dimensions have equal 
radii $R^\prime$, Eq.~(\ref{ADDMPL.EQ}) implies a
particular value of $R^\prime$. However, if it happens that there are two sets of compact extra 
dimensions of unequal size, $\delta$ of them ($\delta \leq n$) with radius $R$, and the other
$(n-\delta)$ with radius $R^\prime$,
then we have in this case, $V_n \sim  {R^\prime}^{(n-\delta)}\,R^{\delta}$. We can in this case
think of $R$ as an independent variable with $R^\prime$ being determined by 
Eq.~(\ref{ADDMPL.EQ}).  

We consider the ADD framework, to which is added three (one for each generation) bulk fermions, 
$\Psi^\alpha (x^\mu,\underline{y})$, that propagate in 4+$\delta$ dimensions 
($\delta$ of them compact with radius $R$), where the indices 
$\alpha$,~$\beta=(1,~2,~3)$ denote the three generations, and $\underline{y}$ 
stands for $\{y^1,...,y^\delta\}$. Since the higher dimensional Lorentz invariance is reduced 
to our 4-dimensional invariance by the presence of the brane (where SM fields are localized), 
we are interested only in keeping track of the 4-dimensional Lorentz transformation 
property of $\Psi$ along the direction of the brane. We consider the situation where the 
SM fields couple only to 2 components of $\Psi$, denoted as $\psi_R(x^\mu,\underline y)$, 
and transforming as a right-handed 2-component Weyl spinor under 4-D Lorentz transformations.
For example, for $\delta = 1$, we denote a 5 dimensional (4 component) spinor as 
\bea
\Psi^\alpha (x^\mu,y) = \pmatrix{\psi_L^\alpha (x^\mu,y) \cr \psi_R^\alpha (x^\mu,y)},
\eea
where the $L$ and $R$ subscripts make explicit the four dimensional Lorentz property, 
and in particular, $\psi^\alpha_{L,R}$ each transforms as a (2 component) Weyl spinor.
In the following, we will keep the formalism general enough to include 
arbitrary number of extra dimensions. 
 
We can split the Lagrangian into a bulk piece and a brane piece,
\bea
{\cal S} &=& \int d^4x~ d^\delta y~\left[{\cal L}_{\rm Bulk}~+~\delta(\underline y)\,
{\cal L}_{Brane}\right].
\eea
${\cal L}_{\rm Bulk}$ contains the Einstein-Hilbert bulk gravity term (which we will not 
show explicitly, but can be found, for example, in Ref.~\cite{Giudice:1998ck}),
the kinetic energy term for the bulk neutrino field $\Psi(x^\mu,\underline{y})$ and in 
general, a bulk Majorana mass term for $\Psi$, which for simplicity we will omit 
(see Ref.~\cite{Dienes:1998sb} for implications of a nonzero bulk Majorana mass).   
${\cal L}_{\rm Brane}$ contains the SM Lagrangian plus an interaction term between SM fields 
and $\psi_R$,
\bea
{\cal L}_{\rm Bulk} &\supset& \bar\Psi^\alpha\, i\Gamma^M D_M \Psi^\alpha, \nonumber \\
{\cal L}_{\rm Brane} &\supset& {\cal L}_{\rm SM} - \left(\frac{\Lambda^\nu_{\alpha\beta}}
{\sqrt{M_*^\delta}}~h\,\psi_R^\beta\,\nu_L^\alpha + h.c. \right),  \label{BULKL.EQ} \\  
{\cal L}_{\rm SM} &\supset& \bar\nu_L^\alpha\, i\gamma^\mu D_\mu \nu_L^\alpha + 
\left(\frac{g}{\sqrt{2}}\,\bar\nu_L^\alpha\, \gamma^\mu e_L^\alpha W_\mu^+ + h.c\right) 
+ ...\, , \nonumber
\eea
where, $\Lambda^\nu_{\alpha\beta}$ is an $O(1)$ Yukawa coupling constant. It should be 
kept in mind that $\psi_R$ is a function of ($x^\mu,\underline y$) whereas the SM fields 
are functions of $x^\mu$ only. The index $M$ runs over \{$x^\mu,\underline y$\}.

We can perform a Kaluza-Klein (KK) expansion of the 4+$\delta$ dimensional theory and obtain 
an equivalent 4-D theory by writing, 
\bea
\psi^\alpha_{R}(x^\mu,\underline{y}) = \sum_{\underline n}~
{\psi_R^\alpha}^{(\underline n)}(x^\mu)~f_{\underline n}(\underline{y}),
\eea
where, $\underline n = (n_1,...,n_\delta)$ is a vector in ``number space'', 
$\psi^{(\underline n)}$ are the KK modes and $f_{\underline n}(\underline{y})$ is a 
complete set over $\underline{y}$. A similar expansion is made for $\psi^\alpha_{L}$. 
To reduce clutter, we will simply write $n$ and $y$ for $\underline n$ and $\underline y$ 
respectively. We will use the notation $n=(0,1,...)$ and $\hat n=(1,...)$ ($\hat n$ excludes 0). 
$f_n$ is an orthonormal set,
\bea
\int_{0}^{2\pi R} d^\delta y ~~ f_n^*(y) f_{m}(y) = \delta^{n m},
\eea 
and a convenient choice is,
\bea
f_{n}(y) = \frac{e^{i \frac{n.y}{R}}}{\sqrt{\frac{S_{\delta-1}}{\delta} R^\delta}} = 
\frac{e^{i \frac{n.y}{R}}}{\sqrt{V_\delta}},
\eea
with $S_{\delta-1}$ the surface ``area'' of a unit sphere in $\delta$ dimensions, and 
$V_\delta \equiv \frac{S_{\delta-1}}{\delta} R^\delta$ is the volume of the extra dimensional space.

We define the fields\footnote{The other linear combinations 
$\left(\psi_R^{\alpha (\hat n)} - \psi_R^{\alpha (-\hat n)} \right)$ and 
$\left(\psi_L^{\alpha (\hat n)} - \psi_L^{\alpha (-\hat n)} \right)$ are decoupled from the SM fields, 
and we will not consider them further. Also, with orbifold compactification, we can project out
$\psi_L^{\alpha (0)}$ so that it is excluded from the particle 
spectrum.}~\cite{Davoudiasl:2002fq},
\bea
\nu^\alpha_R &\equiv& \psi_R^{\alpha (0)}, \nonumber \\
\nu_R^{\alpha (\hat n)} &\equiv& \frac{1}{\sqrt{2}} \left(\psi_R^{\alpha (\hat n)} + 
\psi_R^{\alpha (-\hat n)} \right),  \nonumber \\
\nu_L^{\alpha (\hat n)} &\equiv& \frac{1}{\sqrt{2}} \left(\psi_L^{\alpha (\hat n)} + 
\psi_L^{\alpha (-\hat n)} \right) \ . \nonumber
\eea
We substitute the KK expansion for the bulk fields $\psi_R^\alpha$ and 
$\psi_L^\alpha$ into Eq.~(\ref{BULKL.EQ}) to get the equivalent 4-D theory,
\beq
{\cal L}^{(4)} = {\cal L}_{SM} - \sum_{\alpha=1}^{3}\sum_{\hat n}\left[\frac{|\hat n|}{R}
\left(\nu_R^{\alpha (\hat n)} \nu_L^{\alpha {(\hat n)}} + h.c.\right)\right] -
\sum_{\alpha,\beta=1}^{3} \left[\frac{m_\nu^{\alpha\beta}}{v} \left(h \nu_R^\alpha \nu_L^\beta + 
\sum_{\hat n} \sqrt{2} h \nu_R^{\alpha (\hat n)} \nu_L^\beta \right) + h.c.\right],
\label{EQ4DL1.EQ}
\eeq 
where, $|\hat n|~\equiv~\sqrt{n_1^2+...+n_\delta^2}$. We note here that in ${\cal L}^{(4)}$, there is 
a tower of KK states $(\nu_L^{(\hat n)},\nu_R^{(\hat n)})$ with Dirac masses approximately equal to 
$|\hat n|/R$. Henceforth, we will assume that unless noted otherwise,
repeated generation indices $\alpha,\beta,i,j$, and KK indices $n,\hat n,m$, are summed over.

With $SU(2)$ broken by the Higgs mechanism, 
by the Higgs field acquiring a vacuum expectation value (VEV), $\left<h\right>=v$, we have 
the neutrino masses given by,
\bea
m_\nu^{\alpha \beta} \equiv 
\frac{\Lambda^\nu_{\alpha \beta}~v}{\sqrt{\frac{S_{\delta-1}}{\delta}(M_*R)^\delta}} = 
\frac{\Lambda^\nu_{\alpha \beta}~v}{\sqrt{V_\delta M_*^\delta}},
\eea
and $m_\nu$ can be much smaller than $M_{EW}\sim v$, if $V_\delta$ is large. We show in 
Table~\ref{LAM.TAB} an estimate of the order of magnitude of the $\Lambda^\nu_{\alpha \beta}$
that one needs in order to get neutrino masses of $O(10^{-2})$~eV. The $1/R$ values are chosen 
anticipating the constraint from neutrino oscillation that we will derive in Sec.~\ref{OSCON.SEC}.
We note that $\delta=1$, though the case most studied, requires an unnaturally small $\Lambda$. We will 
therefore include $\delta > 1$ in our analysis.  
\begin{table}
\begin{center}
\caption{Estimate of $\Lambda^\nu_{\alpha \beta}$. ($M_*$ = 1~TeV)}
\begin{tabular}{c|c|c}
$\delta$ & $\frac{1}{R}\ (eV)$ & $\Lambda^\nu_{\alpha \beta}$\\
\hline
\hline
1 & 1 & $10^{-6}$ \\
2 & 1 & 0.1 \\
3 & $10^{3}$ & 1 \\
\end{tabular}
\label{LAM.TAB}
\end{center}
\end{table}

\subsection{Truncation of the KK sum}
\label{KKT.SEC}
Though $\hat n_i$ can in principle go up to $\infty$, leading to an infinite tower of Dirac states 
with masses $|\hat n|/R$, we take the view that this extra dimensional field theory description 
is valid up to the cutoff scale $M_*$, and therefore truncate the $\hat n_i$ such that the 
highest KK mass is $M_*$. We define $N$ to be $N/R=M_*$. For $\delta > 1$, the state with mass 
$\frac{|\hat n|}{R}$ can be degenerate, and we denote the degeneracy at the $\hat n^{\rm th}$ 
level by $d_{\hat n}$. (Strictly speaking, we should denote this as $d_{|\hat n|}$, but we will 
just write this as $d_{\hat n}$.) For example, for $\delta=3$, the state with mass 
$1/R$ has $d_1 = 3$, corresponding to 
$(\hat n_1,\hat n_2,\hat n_3)\rightarrow (1,0,0),\ (0,1,0)\ {\rm and}\ (0,0,1)$, all
of which have the same mass. For large $|\hat n|$, the leading power dependence of $d_{\hat n}$
in $\delta$ extra dimensions is given by $d_{\hat n} = c_{\hat n} |\hat n|^{\delta - 1}$, 
where the $c_{\hat n}$ are $O(1)$ numbers. We define $d_0 \equiv 1$.

For large $|\hat n|$, we can think of the $\hat n_i$ as a continuum and the leading behavior 
is given by the surface of the $(\delta-1)$-sphere of radius $|\hat n|$ in number space, 
\bea
d_{\hat n} \sim S_{\delta-1} |\hat n|^{\delta-1} \ \ \ \ \ \ {\rm (in\ \delta\ dimensions).} 
\label{DNC.EQ} \\
d^{\delta = 1}_{\hat n} = 1, \ \ \ \ \ \ d^{\delta = 2}_{\hat n} \sim 2\pi\hat n, \ \ \ \ \ \  
d^{\delta = 3}_{\hat n} \sim 4\pi\hat n^2 \ . \nonumber
\eea
For example, for $\delta=3$, $d_{\hat n} \sim 4\pi |\hat n|^2$, which is the surface of the 2-sphere with 
radius $|\hat n|$. Thus, $N$ is the radius of the biggest sphere in $\{n_i\}$ space such that
$N/R = M_*$. We will often use the continuum approximation for estimating various quantities.

The sum over the KK states of certain quantities can be divergent and can depend on $N$. 
We will elaborate more on this later, and we will see that the 
probability of active neutrinos oscillating into heavier sterile KK states can depend on N, 
especially strongly for $\delta>3$.

\subsection{Mass Matrix Diagonalization}
\label{MMD.SEC}
At the outset, we can make the Yukawa coupling in Eq.~(\ref{EQ4DL1.EQ}) diagonal in generation space 
($\alpha, \beta$) with the rotations~\cite{Davoudiasl:2002fq},
\bea
\nu_L^\alpha = l^{\alpha i} \nu_L^{\prime i} &,& \ \ \
\nu_R^\alpha = (r^{\alpha i})^* \nu_R^{\prime i} \ , \nonumber \\
\nu_R^{\alpha (\hat n)} = (r^{\alpha i})^* \nu_R^{\prime i (\hat n)} &,& \ \ \
\nu_L^{\alpha (\hat n)} = r^{\alpha i} \nu_L^{\prime i (\hat n)} \ , \nonumber \\
e_L^\alpha = l_e^{\alpha i} e_L^{\prime i} &,& \ \ \
e_R^\alpha = (r_e^{\alpha i})^* e_R^{\prime i} \ , \nonumber
\label{MNSROT.EQ}
\eea
 where the unitary matrices $l$ and $r$ are chosen to diagonalize $m_\nu^{\alpha\beta}$, so that, 
$(r^{\alpha i})^*~m_\nu^{\alpha\beta}~(l^{\beta j})=m_\nu^{i} \delta^{i j}$. Similarly, $l_e$ 
and $r_e$ are chosen to diagonalize the electron-type mass matrix. After these 
rotations, Eq.~(\ref{EQ4DL1.EQ}) becomes,
\beq
{\cal L}^{(4)} = {\cal L}_{SM} -
\frac{|\hat n|}{R}\left(\nu_R^{\prime i (\hat n)} \nu_L^{\prime i (\hat n)} + h.c.\right) -
\left[\frac{m_\nu^i}{v} \left(h \nu_R^{\prime i} \nu_L^{\prime i} + 
\sum_{\hat n} \sqrt{2} h \nu_R^{\prime i (\hat n)} \nu_L^{\prime i} \right)
  + h.c.\right] \ .
\label{EQ4DLNG.EQ}
\eeq 
In the usual way, the charged current interactions now become proportional to the MNS 
matrix~\cite{Maki:mu},
\beq
V_{MNS} \equiv l_e^\dag\, l \ .
\label{MNSDEF.EQ}
\eeq

By defining the Dirac spinors,
\beq
\nu^{\prime i} \equiv \pmatrix{\nu_L^{\prime i} \cr \nu_R^{\prime i}},~~\nu^{\prime i(1)} \equiv \pmatrix{\nu_L^{\prime i(1)} \cr \nu_R^{\prime i(1)}}
~.~.~.~\nu^{\prime i(n)} \equiv \pmatrix{\nu_L^{\prime i(n)} \cr \nu_R^{\prime i(n)}},
\eeq
we can rewrite Eq.~(\ref{EQ4DLNG.EQ}) as, 
\beq
{\cal L}^{(4)} = {\cal L}_{SM} -
\frac{|\hat n|}{R} \bar{\nu^{\prime i(\hat n)}} \nu^{\prime i(\hat n)} -
\frac{m_\nu^i}{v} \left[h \bar{\nu^{\prime i}} \nu^{\prime i} + 
\sqrt{2} h \left(\sum_{\hat n} \bar{\nu^{\prime i(\hat n)}} P_L \nu^{\prime i} + h.c.\right) \right],
\label{EQ4DLDAPP.EQ}
\eeq
where $P_{L(R)}\equiv \left( 1\mp \gamma_5\right)$.

To reduce clutter, in the rest of this section, we will suppress writing the generation 
index $i$ and the prime on the fields. Thus for each $i=1,2,3$, the mass matrix in KK space 
can be diagonalized independently using the procedure described next. 

For each $i$, the neutrino mass term obtained by setting 
$\left<h\right>=v$ is,
\beq
{\cal L}^{(4)}_{\rm mass} =  \bar{\nu_D}~{\cal M}_D~\nu_D , 
\eeq
with $(\nu_D)^T = (\nu~\nu^{(1)}~...~\nu^{(\hat n)}~...)$ and the mass matrix is given by,
\beq
{\cal M}_D = \pmatrix{
m_\nu & \sqrt{2}m_\nu P_R & . & . & . & \sqrt{2}m_\nu P_R & . & . & . \cr
\sqrt{2}m_\nu P_L & \frac{1}{R} &  &  &  &  &  &  & \cr
. &  & . &  &  &  &  &  &  \cr
. &  &  & . &  &  &  &  &  \cr
. &  &  &  & . &  &  &  &  \cr
\sqrt{2}m_\nu P_L &  &  &  &  & (\frac{|\hat n|}{R})_{d_{\hat n}\times d_{\hat n}} &  &  &  \cr
. &  &  &  &  &  & . &  &  \cr
. &  &  &  &  &  &  & . &  \cr
. &  &  &  &  &  &  &  & \frac{|N|}{R} \cr
}.
\label{NUMD.EQ}
\eeq
We keep in mind that the ${\hat n}^{\rm th}$ state is degenerate with degeneracy $d_{\hat n}$.

We define 
\bea
\xi \equiv m_\nu R, \ \ \ \ \  
\xi_{\hat n} \equiv \frac{\xi}{|\hat n|}.  
\eea
We will restrict ourselves to the situation when $\sum_{\hat n} \xi_{\hat n}^2 d_{\hat n} \ll 1$ 
(for all $\delta$), 
since, as we will show later, this is the condition implied by experimental data on oscillation 
into sterile states. 

We can diagonalize ${\cal M}_D$ by two unitary matrices $L$ and $R$ so that
\beq
\nu_D = \left(L\, P_L + R\, P_R \right) \tilde{\nu}_D,
\eeq
where $\tilde{\nu}_D$ is the mass eigenvector. We perform the diagonalization perturbatively, and to $O(\xi^2)$, the $L$ and $R$ are, 
\bea
L &=& \pmatrix{
(1-\sum \xi_{\hat n}^2 d_{\hat n}) & \sqrt{2}\xi_1 & . & . & . & \sqrt{2}\xi_{\hat n}  & . & . & \sqrt{2}\xi_N \cr
-\sqrt{2}\xi_1 & (1-\xi_1^2) &  &  &  &  &  &  &  \cr
. &  & . &  &  &  &  &  &  \cr
. &  &  & . &  &  &  &  &  \cr
. &  &  &  & . &  &  &  &  \cr
-\sqrt{2}\xi_{\hat n} &  &  &  &  & (1-\xi_{\hat n}^2)_{d_{\hat n}\times d_{\hat n}} &  &  &  \cr
. &  &  &  &  &  & . &  &  \cr
. &  &  &  &  &  &  & . &  \cr
-\sqrt{2}\xi_N &  &  &  &  &  &  &  & (1-\xi_N^2) \cr
}, 
\label{DL.EQ} \\
\nonumber \\
R &=& \pmatrix{
1 & \sqrt{2}\xi_1^2 & . & . & . & \sqrt{2}\xi_{\hat n}^2  & . & . & \sqrt{2}\xi_N^2 \cr
-\sqrt{2}\xi_1^2 & 1 &  &  &  &  &  &  &  \cr
. &  & . &  &  &  &  &  &  \cr
. &  &  & . &  &  &  &  &  \cr
. &  &  &  & . &  &  &  &  \cr
-\sqrt{2}\xi_{\hat n}^2 &  &  &  &  & (1)_{d_{\hat n}\times d_{\hat n}} &  &  &  \cr
. &  &  &  &  &  & . &  &  \cr
. &  &  &  &  &  &  & . &  \cr
-\sqrt{2}\xi_N^2 &  &  &  &  &  &  &  & 1 \cr
}.
\eea

The mass eigenvalues are the diagonal elements of $R^\dag {\cal M}_D L$. (See Appendix~\ref{MDDS.APP} 
for an alternate approach to diagonalization and for details on obtaining the mass eigenvalues.)
We find the lowest mass eigenvalue, 
\beq
m^{(0)} \approx  m_\nu \left(1-\xi^2 \sum_{\hat n} \frac{d_{\hat n}}{|\hat n|^2}\right).
\label{M0.EQ}
\eeq
Given that we take $\xi$ to be small enough so that $\sum_{\hat n} \xi_{\hat n}^2 d_{\hat n} \ll 1$, we have 
$m^{(0)}\approx m_\nu$. We note that henceforth, when we write $m$, we mean $m^{(0)}$.  
The heavier mass eigenvalues $m^{(\hat n)}$ at the $\hat n^{\rm th}$ level, with $d_{\hat n}$ states, are 
\beq
m^{(\hat n)} =
\left\{
\begin{array}{l} %
\displaystyle \frac{|\hat n|}{R}\left[1+(\xi^2+\xi^4)d_{\hat n}\right] \ \ \ \ \ {\rm for\ 1\ state},\\
\nonumber \\ [-5mm]
\displaystyle \frac{|\hat n|}{R} \ \ \ \ \ \ \ \ \ \ \ {\rm for\ the\ other\ (d_{\hat n}-1)\ states}.
\end{array}
\right.
\label{MN.EQ}
\eeq

In summary, reintroducing the generation index $i$ and the KK index 
$n\rightarrow \{0,\hat n\};~\hat n=1,...,N$, we can write the flavor state $\nu_L^\alpha$ in 
terms of the mass eigenstates $\tilde\nu_L^{i (n)}$ as
\beq
\nu_L^\alpha = l^{\alpha i} L_i^{0n}\tilde\nu_L^{i (n)}.  
\label{FMN.EQ}
\eeq
We can write a similar expression for the right-handed neutrino, but we will not need that.

\section{Neutrino Oscillation Constraints}
\label{OSCON.SEC}

Neutrino oscillation experiments measure the probability of producing an active neutrino flavor 
$\alpha$, and detecting an active flavor $\beta$ a distance $L$ away.
Recent experiments, particularly Super-Kamiokande and SNO, indicate that 
the oscillation among the three active species $\nu_e,~\nu_\mu,~\nu_\tau$ is the best fit to 
data. We will denote the probability of an active species $\nu_\alpha$ oscillating into 
another active species $\nu_\beta$ after a distance L as $P_{\nu_\alpha \rightarrow \nu_\beta}$. 
The mixing of an active species into a sterile species $\nu_s$, 
$P_{\nu_\alpha\rightarrow \nu_s}\equiv 1-\sum_{\beta}P_{\nu_\alpha\rightarrow \nu_\beta}$, 
is furthermore, strongly constrained by CHOOZ~\cite{Apollonio:1999ae} and by fits to the atmospheric 
oscillation data. We will thus assume that the standard three active species 
oscillation with the usual MNS matrix provides the best fit to data, and the oscillation
into the heavier KK states $\nu^{(\hat n)}$ (sterile states), would have to be small enough not 
to violate the constraints on $P_{\nu_\alpha\rightarrow \nu_s}$. This was the view 
taken also in Ref.~\cite{Davoudiasl:2002fq}, where it is noted that the constraints on the 
mixing into sterile states are the following (at the 90\% C.L.):
\bea
{\rm CHOOZ} &\qquad&  P_{\nu_e\rightarrow \nu_s} < 0.058  , \label{CHLIM.EQ} \\
{\rm Atmospheric\ \nu} &\qquad& 
\left\{
\begin{array}{l}
P_{\nu_\mu\rightarrow \nu_s} - P_{\nu_e\rightarrow \nu_s} < 0.17 \ , \\
\frac{1}{2}\left[ P_{\nu_\mu\rightarrow \nu_s} + P_{\nu_e\rightarrow \nu_s}\right] < 0.39 \ . 
\end{array} 
\right.
\label{ATMLIM.EQ}
\eea
We have chosen these bounds as they place the strongest constraints on our model. 

The probability for an active species known to be $\nu_\alpha$ at $t=0$, to oscillate into the active
species $\nu_\beta$ after time $t$ (after traveling a length $L$), is determined by the Hamiltonian $H$,
and is given by the time evolution operator $e^{-iHt}$. (For a concise account on neutrino oscillation,
 see for example Ref.~\cite{Kayser:1981ye}.) 
Therefore,
\beq
P_{\nu_\alpha \rightarrow \nu_\beta} = \left| \left<\nu_L^\beta | e^{-iHt} | \nu_L^\alpha\right> \right|^2 \ .
\eeq
Using Eq.~(\ref{FMN.EQ}) to write $\nu_L^\alpha$ in terms of the energy eigenstates 
$\tilde\nu_L^{i (n)}$, and since 
\beq
H~\tilde\nu_L^{i (n)} = E^{i (n)}~\tilde\nu_L^{i (n)}, \nonumber
\eeq
we get,
\beq
P_{\nu_\alpha \rightarrow \nu_\beta} = \left|
 l^{\beta i *} l^{\alpha i} |L_i^{0n}|^2 d_n  e^{-i E^{i (n)} L} 
\right|^2 \ , 
\eeq
where $L_i^{0n}$ is the first row in Eq.~(\ref{DL.EQ}) and $d_0 = 1$.
For a neutrino beam with energy $E_\nu$ and momentum $p_\nu$, in the 
relativistic limit, we have, 
$E^{i (n)} \approx p_\nu + m_i^{(n)\, 2}/(2 E_\nu)$, and therefore,
\beq
P_{\nu_\alpha \rightarrow \nu_\beta} = \left| l^{\beta i *} l^{\alpha i} 
\left( |L_i^{00}|^2 e^{-i \frac{L}{2 E_\nu} m_i^2}  + |L_i^{0\hat n}|^2 d_{\hat n}  e^{-i \frac{L}{2 E_\nu}m_i^{(\hat n)\, 2}}
 \right)  \right|^2 \ .
\eeq
$m^{i (n)}$, the mass eigenvalues, are as given in Eqs.~(\ref{M0.EQ})~and~(\ref{MN.EQ}). 
From Eq.~(\ref{DL.EQ}) we have
\beq
L_i^{00} = 1-\sum_{\hat n} \frac{\xi_i^2}{\hat n^2} d_{\hat n} \ ,\ \ \ \  
L_i^{0\hat n} = \sqrt{2}\frac{\xi_i}{\hat n} \nonumber 
\eeq 
which implies to $O(\xi^2)$
\beq
\sum_{\beta} P_{\nu_\alpha \rightarrow \nu_\beta} = 1 - 8 |l^{\alpha i}|^2 \xi_i^2 \sum_{\hat n} \frac{d_{\hat n}}{\hat n^2} \sin^2\left( \frac{L\hat n^2}{4 E_\nu R^2} \right) \ ,
\eeq
where we assume that $\frac{\hat n}{R} \gg m_i$, which we will see soon is required to
satisfy the experimental constraints. Hence, the probability for an active state to oscillate into 
sterile states is
\beq
P_{\nu_\alpha \rightarrow \nu_s} = 1-\sum_{\beta} P_{\nu_\alpha \rightarrow \nu_\beta} = 8 |l^{\alpha i}|^2 \xi_i^2 \sum_{\hat n} \frac{d_{\hat n}}{\hat n^2} \sin^2\left( \frac{L\hat n^2}{4 E_\nu R^2} \right) \ .
\label{PAS.EQ}
\eeq

Before we can use Eq.~(\ref{PAS.EQ}) to compute the bounds on $1/R$, we have to specify the $m_i$ and
the $l^{\alpha i}$. 
We take the $m_i$ given by Eq.~(\ref{M0.EQ}) to be consistent with the mass differences extracted 
from standard fits to the solar and atmospheric neutrino oscillation data given in 
Eq.~(\ref{DMSQ.EQ}). So far we have precise information only on the mass differences from the 
oscillation data, with upper limits on the masses otherwise. The limits are 
$(m_{\nu_e}, m_{\nu_\mu}, m_{\nu_\tau}) < (3~{\rm eV}, 0.19~{\rm MeV}, 18.2~{\rm MeV})$
~\cite{Hagiwara:fs}, from final state lepton spectrum,
and, the sum of the neutrino masses $\sum_j m_{\nu_j} < {\rm few}$~eV, from cosmology 
(for reviews, see Ref.~\cite{Dolgov:2002wy}).  
It is usual to consider the following three mass schemes, all of which are consistent with 
Eq.~(\ref{DMSQ.EQ}):
\begin{itemize}
\item[(i)] Normal Hierarchy: $m_1\approx 0$, $m_2\approx 0.008$~eV and $m_3\approx 0.05$~eV. 
\item[(ii)] Inverted Hierarchy: $m_1,m_2\approx 0.05$~eV and $m_3\approx 0$. 
\item[(iii)] Degenerate: $m_1,m_2,m_3$ all at some mass scale less than the limits given above. 
Here, to illustrate
the character of the bounds, we take the three masses to be around 1~eV, an arbitrary choice. 
\end{itemize}
We can work in the basis in which the charged lepton mass matrix is diagonal. In this case we have from
Eq.~(\ref{MNSDEF.EQ}), $V_{MNS} = l$. The bounds we 
derive are not very sensitive to the precise values chosen for $l$. Based on the $V_{MNS}$ obtained 
from a global fit to the oscillation data, we take the $l$ to be such that the solar neutrino 
mixing angle between $\nu_e~\&~\nu_\mu$ is given by $\tan^2\theta_{e\mu}=0.4$, and the 
atmospheric oscillation mixing between $\nu_\mu~\&~\nu_\tau $ is maximal. This implies
\beq
l = \pmatrix{
0.847 & 0.531 & 0 \cr
-0.376 & 0.599 & 0.707\cr
0.375 & -0.599 & 0.707
}.
\eeq

Given a particular mass scheme, we can then derive the bounds on $1/R$ using Eq.~(\ref{PAS.EQ}). In this
equation we note that for large KK masses, the argument of the sine function 
is large, leading to a rapid oscillation with distance. We therefore use
the average value $\left<\sin^2\left( \frac{L\hat n^2}{4 E R^2}\right)\right> = 1/2$, and require that the 
experimental constraint be satisfied by
\beq
\left<P_{\nu_\alpha \rightarrow \nu_s}\right> = 4 |l^{\alpha i}|^2 \xi_i^2 \sum_{\hat n} \frac{d_{\hat n}}{\hat n^2} \ .
\label{PASAV.EQ}
\eeq
We remind the reader that in the continuum approximation, the $d_{\hat n}$ is given in Eq.~(\ref{DNC.EQ}).
The bounds on $1/R$ obtained by requiring Eq.~(\ref{PASAV.EQ}) to satisfy the limits in 
Eqs.~(\ref{CHLIM.EQ})~and~(\ref{ATMLIM.EQ})  are summarized in Table~\ref{OBND.TAB}. For the lighter KK states,
with $1/R \sim 1$~eV, as could be the case for $\delta=1,2$, the argument of the sine function in 
Eq.~(\ref{PAS.EQ}) may not be large and a numerical calculation may be necessary. However, since the number
of such states are small, we expect that the bounds we found using the average value is a good approximation. 
Ref.~\cite{Davoudiasl:2002fq} has presented the bounds for $\delta=1$ by performing an exact 
numerical calculation, and we find that our bounds are quite similar to theirs.

\begin{table}
\begin{center}
\caption{Lower bound on $\frac{1}{R}$~(eV) for right-handed neutrinos in $\delta=1,2,3$ 
extra dimensions.}
\begin{tabular}{|c|c|c|c|c|c|c|}
\hline 
&
\multicolumn{2}{c|}{Normal}&
\multicolumn{2}{c|}{Inverted}&
\multicolumn{2}{c|}{Degenerate\ ($m \approx 1$~eV)}
\tabularnewline
 \cline{2-7} 
&
CHOOZ&
Atm&
CHOOZ&
Atm&
CHOOZ&
Atm\tabularnewline
\hline
\hline 
$\delta=1$&
0.03 & 0.15 & 0.5 & 0.13 & 10.6 & 4.1
\tabularnewline
\hline 
$\delta=2$&
0.32 & 1.5 & 5.3 & 1.3 & 100 & 41.7
\tabularnewline
\hline 
$\delta=3$&
$2.4\times 10^3$ & $5.6\times 10^3$ & $1.2\times 10^4$ & $4.9\times 10^3$ & $10^5$ & $5\times 10^4$ 
\tabularnewline
\hline
\end{tabular}
\label{OBND.TAB}
\end{center}
\end{table}

We would like to stress here that for $\delta > 1$, the sum over the KK states, for example in 
Eq.~(\ref{PASAV.EQ}), is not convergent, given that the degeneracy at the 
$n^{\rm th}$ level is $d_n \sim n^{\delta-1}$. To illustrate this further, considering 
$\delta=3$ with $d_n\sim 4\pi n^2$, we get equal probability, for all $n$, for the active neutrino 
to mix to the $n^{\rm th}$ KK level, after including the degeneracy. Therefore our estimates are 
valid only up to $O(1)$ coefficients since we are sensitive to the manner in which our four dimensional 
KK description is completed to a more fundamental theory at the scale $M_*$, the cutoff of our 
theory.\footnote{For a related discussion in the graviton sector, see Ref.~\cite{Bando:1999di}.} 
An additional source of
uncertainty comes from using the continuum approximation rather than performing the discrete sum
over $\hat n$.

One might wonder if in a certain high energy completion (or if one includes higher dimension 
operators in the effective theory), it is possible that the off-diagonal entries in the 
neutrino mass matrix, Eq.~(\ref{NUMD.EQ}), could be vanishingly small compared to the 00 entry, 
so that the above mentioned oscillation constraints could be severely weakened. 
Since it is the same operator that generates the 00 and the off-diagonal entries, 
c.f. Eqs.~(\ref{BULKL.EQ})~and~(\ref{EQ4DL1.EQ}), this would mean that the 00 term also has to be 
vanishingly small, which would conflict with the experimental constraint in Eq.~(\ref{DMSQ.EQ}). 
We thus conclude that the 
oscillation bounds on $1/R$ that we estimated above are general to the class of theories, 
in which only right-handed neutrinos can propagate in bulk,  
 though they depend to some level on how the theory is completed at the scale $M_*$. 
For $\delta>3$ the active state would oscillate mostly to the heaviest states and we would not be able
to reliably estimate the oscillation probability using the effective theory that we are working with. 
We thus restrict out conclusions to $\delta \leq 3$. 

\section{Unitarity Constraints}
\label{UNICON.SEC}
We would like our extra dimensional theory to be unitary at each order in perturbation in order 
to have the ability to make reliable predictions. The amplitude for certain processes that 
receive contributions from a tower of KK intermediate states can grow with the center-of-mass (c.m.)
collision energy $\sqrt{s}$. For such processes, the requirement of perturbative 
unitarity can be particularly stringent as $\sqrt{s}$ increases. 

For instance, the number of KK states accessible at a 
collision energy $\sqrt{s}$ is of the order of $(\sqrt{s} R)^\delta$, which can be large.
For the amplitude of a scattering process that grows with the number of KK 
states in the intermediate state, requiring 
that unitarity be maintained for $\sqrt{s}\leq M_*$ can lead to a lower bound on $1/R$. 
As pointed out in Sec.~\ref{KKT.SEC}, the sum over the KK states can be divergent, and 
therefore, we cut the sum off at $N_s = \sqrt{s} R$.  

\begin{figure}
\begin{center}
\scalebox{0.50}{\includegraphics{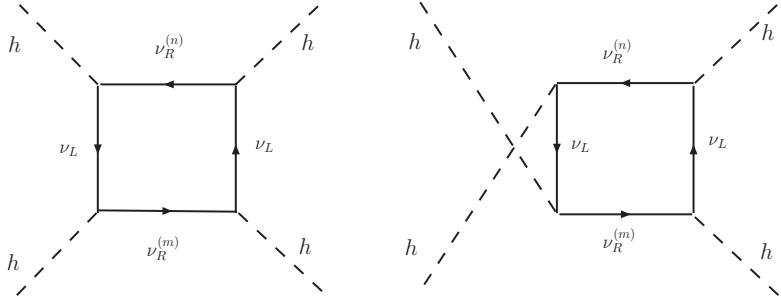}}
\caption{Diagrams that dominantly contribute to ${\rm Im}({\cal A})$. \label{fig:unitarity_diag}}
\end{center}
\end{figure}

We consider the imagined scattering of a pair of Higgs bosons 
$hh \rightarrow hh$, as shown in Fig.~\ref{fig:unitarity_diag}, to study the unitarity 
constraints on $1/R$ as a function of $\sqrt{s}$. 
The two incoming (outgoing) momenta of the
Higgs bosons are labeled by $p_1$ and $p_2$ ($p_3$ and $p_4$). With $p_1+p_2=p_3+p_4$, 
we define the Mandelstam variables in the usual way, $s=2p_1\cdot p_2 + 2 m_h^2$, 
$t=-2p_1\cdot p_4+2m_h^2$ and $u=-s-t+4m_h^2=-2p_1\cdot p_3+2m_h^2$. 
The kinematics of $2\rightarrow2$ scattering are simply described by $s$ and one scattering angle, 
$\theta$, in terms of which $\displaystyle t=-\frac{s}{2} \left(1-{4m_h^2}/{s}\right)(1-\cos{\theta})$ and 
$\displaystyle u=-\frac{s}{2} \left(1-{4m_h^2}/{s}\right)(1+\cos{\theta})$. 
The amplitude ${\cal A}$ can be expanded in Legendre polynomials as
\bea
{\cal A}(h h \rightarrow h h)=(32\pi) \sum^{\infty}_{l=0}(2l+1)P_l(\cos\theta)\, a_l \ .
\label{PWE.EQ}
\eea
Given the scattering amplitude ${\cal A}$, the partial wave amplitude $a_l$ are determined by
\bea
a_l=\frac{1}{64\pi}\int^{1}_{-1} \,{\rm d}\cos\theta \,\,P_l(\cos\theta)\,\, 
{\cal A}(h h \rightarrow h h) \ . 
\label{eq:partial_amp}
\eea
For the process we are considering, for unitarity to be satisfied, it is sufficient that 
\bea
{\rm Im}({a_0}) < 1 \ , 
\label{eq:partial_bounds}
\eea
where $a_0$ is the ${\rm S}$-wave amplitude. We will thus focus on calculating ${\rm Im}({\cal A})$ in
order to apply this bound. 

The dominant contribution to the imaginary part of the scattering amplitude $\mathcal A$ 
can be written as 
\beq
{\rm Im}\left({\cal A}\right) = {\rm Im}\left(\sum_{nm} 
{\mathcal M^{n m}} d_{n} d_{m} \right)
\label{IMNM.EQ}
\eeq
where the Feynman graphs for ${\mathcal M^{nm}}$ are shown in Fig.~\ref{fig:unitarity_diag}, in which
we have shown only the diagrams that contribute dominantly to ${\rm Im}({\cal A})$. 

Before detailing the full calculation, we present an order of magnitude estimate of ${\rm Im}({\cal A})$.
Due to the unitarity of the ${\rm S}$-matrix, ${\rm Im}({\cal A})$ is given by the sum of
the cut diagrams. We obtain an estimate by ignoring the KK masses in the intermediate state, and furthermore, 
by using the continuum approximation, c.f. Eq.~(\ref{DNC.EQ}). We define $N_s$ so that the
highest KK state accessible to the cut diagram is given by $\frac{N_s}{R}=\sqrt{s}$, and we get
\bea
{\rm Im}({\cal A}) &\sim & \frac{1}{16\pi} \sum_{nm}^{N_s} \left(\frac{m_\nu}{v}\right)^4 \left(S_{\delta-1} 
|n|^{\delta-1}\right) \left(S_{\delta-1} |m|^{\delta-1}\right) \ , \nonumber \\
 & \sim & \frac{1}{16\pi} \left(\frac{m_\nu}{v}\right)^4 \frac{1}{2}\; 
    \left(\frac{S_{\delta-1}}{\delta}\right)^2 N_s^{2\delta} \nonumber \\ 
 & \sim & \frac{1}{32\pi} \left(\frac{m_\nu}{v}\right)^4 
  \left(\frac{S_{\delta-1}}{\delta}\right)^2 (\sqrt{s} R)^{2\delta}    \ . 
\label{AEST.EQ}
\eea
We see from this estimate that the amplitude can grow quite steeply, especially for larger $\delta$,
and the suppression due to the small coupling $m_\nu/v$ can be overcome by the growth in the number of states. 
Thus, for the theory to be unitary up to $\sqrt{s}\approx M_*$, Eq.~(\ref{eq:partial_bounds}) implies
a lower bound on $1/R$. 

We have performed a detailed calculation by taking into account the KK masses. We perform a change
of variables to write $n$ in terms of $m^{(n)}$, c.f. Eqs.~(\ref{M0.EQ})~and~(\ref{MN.EQ}) and use the 
continuum approximation to convert the double sum in Eq.~(\ref{IMNM.EQ}) into a double integral to get   
\bea
{\rm Im}({\mathcal A})=\frac{1}{16\pi} S^{2}_{\delta-1} R^{2\delta}
\int^{\sqrt{s}}_0\!{\rm d}m^{(n)}\int^{\sqrt{s}-m^{(n)}}_0\!{\rm d}m^{(m)} 
\left[ m^{(n)} {m^{(m)}}  \right]^{\delta-1} {\rm Im}({\mathcal M}^{(nm)}) .
\label{ANM.EQ}
\eea
We use FeynCalc~\cite{Feyncalc} to calculate ${\mathcal M}^{(nm)}$, and we get
\bea
{\mathcal M^{(nm)}}&=&\left(\frac{m_{\nu}}{v}\right)^4
\pi^2 \Biggl{\{} 4 B_0(s,\mones,\mtwos)+2 B_0(t,\mns,\mns)+2B_0(u,\mns,\mns) \\ \nonumber
&&~~~~+f^c_1(s) \,\, C_0(\mhs,\mhs,s,\mones,\mns,\mtwos) \\ \nonumber
&&~~~~+\biggl[f^c_2(t)\,\, C_0(\mhs,\mhs,t,\mns,\mones,\mns)+(t\rightarrow u)\biggl] \\ \nonumber
&&~~~~+\biggl[f^c_3(t)\,\, C_0(\mns,\mhs,t,\mns,\mtwos,\mns)+(t\rightarrow u)\biggl] \\ \nonumber
&&~~~~+\biggl[f^d(t)\,\,D_0(\mhs,\mhs,\mhs,\mhs,s,t,\mones,\mns,\mtwos,\mns)+(t\rightarrow u) \biggl]\Biggl{\}},
\eea
where $B_0$, $C_0$ and $D_0$ are the Passarino-Veltman (PV) scalar functions~\cite{Passarino:1978jh}, and
$f^c_1$, $f^c_2$, $f^c_3$ and $f^d$ are defined as 
\bea
f^c_1(s)&=&4\left(s+2\mns+2\mone\mn+2\mtwo\mn+2\mone\mtwo-2\mhs\right),\\ \nonumber
f^c_2(t)&=&t+2\mones +4\mn\mone+2\mns-2\mhs,\\ \nonumber
f^c_3(t)&=&t+2\mtwos +\,4\mn\mtwo+2\mns-2\mhs,\\ \nonumber
f^d(t)&=& t\left((\mone-\mtwo)^2-s\right)
         +2\left((\mone+\mn)^2-\mhs\right)\left((\mtwo+\mn)^2-\mhs\right).
\eea
\begin{figure}
\begin{center}
\scalebox{0.60}{\includegraphics{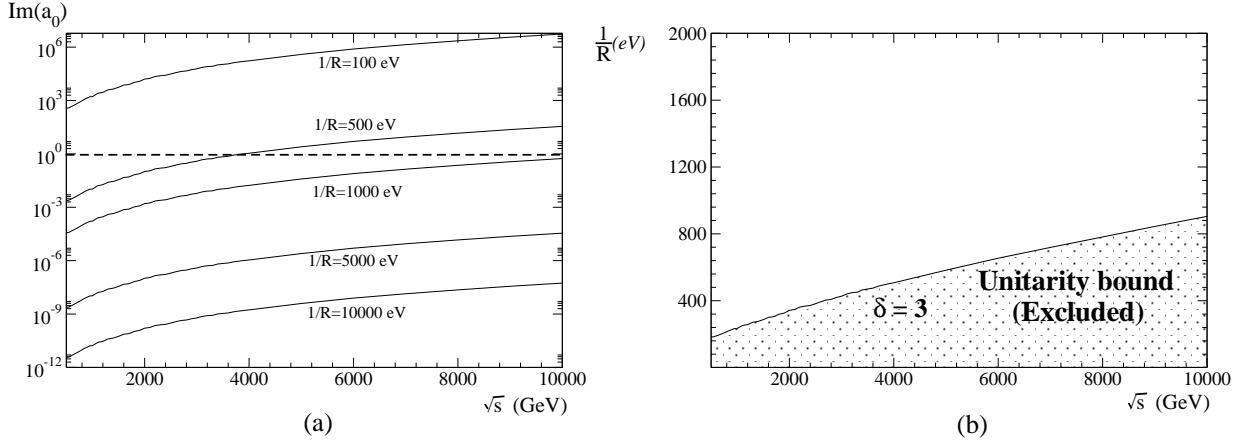}}
\caption{Unitarity bounds for $\delta=3$. The theory is unitary if the solid line in (a) corresponding 
to the given $1/R$ does not cross the dotted line (${\rm Im}(a_0)=1$) at a $\sqrt{s}$ less than $M_*$.
The unitarity constraint on the 1/R vs. $\sqrt{s}$ plane is shown in (b). 
\label{fig:unitarity_bound}}
\end{center}
\end{figure}
The functions $f^c_1$, $f^c_2$, $f^c_3$ and $f^d$ are all real, and therefore, 
${\rm Im}({\mathcal M}^{(nm)})$ is given by the imaginary parts of the PV scalar functions. 
We present the calculation of the imaginary parts of the PV functions $B_0$, $C_0$, $D_0$, 
in Appendix~\ref{IPV.APP}. To obtain ${\rm Im}({\mathcal A})$, we then perform a Monte-Carlo 
integration of Eq.~(\ref{ANM.EQ}) over $m^{(n)}$ and $m^{(m)}$, and expand it into partial waves 
using Eq.~(\ref{PWE.EQ}). We  show the resulting ${\rm Im}(a_0)$ in Fig.~\ref{fig:unitarity_bound}(a) 
for $\delta=3$, as a function of $\sqrt{s}$ for different choices of $1/R$ (from $10^2$~eV to $10^4$~eV). 
For $\delta=3$, the region in $1/R$ vs. $\sqrt{s}$ that satisfies the unitarity bound, c.f. 
Eq.~(\ref{eq:partial_bounds}), is shown in Fig.~\ref{fig:unitarity_bound}(b). We infer from these
plots that for $\delta=3$ and $\sqrt{s}=1$~TeV, we require $1/R > 250$~eV, for unitarity not to be 
violated. 

Using the bound just obtained for $\delta=3$, we can use Eq.~(\ref{AEST.EQ}) to estimate the bounds
for $\delta=1,2$.  For $\delta=2$ and $\sqrt{s}=1$~TeV, we find the unitarity bound, $1/R > 10^{-3}$~eV.  
For $\delta = 1$, we find that the unitarity bound is very weak. 

\section{Conclusions}
\label{CONCL.SEC}

The Standard Model (SM) of high energy physics suffers from the gauge hierarchy problem, 
which is the fine tuning required to maintain a low electroweak scale ($M_{EW} \sim 10^3$~GeV)
in the presence of another seemingly fundamental scale, the Planck scale (the scale of gravity, 
$M_{pl}\sim 10^{19}$~GeV). Theories with $n$ large extra dimensions have the potential to 
address the hierarchy problem as shown by Arkani-Hamed, Dimopoulos and Dvali 
(ADD)~\cite{Arkani-Hamed:1998rs,Antoniadis:1998ig}. 

Recent neutrino oscillation experiments~\cite{Fukuda:1998mi,Ahmad:2002ka,Wolfenstein:1977ue,Bahcall:2002hv} 
have suggested a small non-zero mass for the neutrino. 
It has been shown~\cite{Dienes:1998sb,Arkani-Hamed:1998vp,Dvali:1999cn} that if right-handed bulk 
neutrinos (SM gauge singlets) 
propagate in $\delta$ of these large extra dimensions ($\delta \leq n$), then the 
smallness of the neutrino mass can be explained. 

We considered such bulk neutrinos in the ADD framework where the fundamental scale in nature is postulated to
be $M_* \sim 1$~TeV. Theories with $\delta \geq 2$ naturally give the right order of neutrino masses
with the coefficient of the Yukawa interaction for the neutrino being $O(1)$. However, for 
$\delta = 1$, with the requirement that oscillation constraints be satisfied, this coefficient has 
to be $O(10^{-5})$, c.f. Table~\ref{LAM.TAB}, making the theory unnatural. 

We performed the Kaluza-Klein~(KK) expansion of the bulk neutrino field to obtain the equivalent 
four dimensional theory starting from a $4+\delta$ dimensional theory. It is common in the 
literature to analyze theories with $\delta = 1,2$, while in this paper we considered 
$\delta = 1,2,3$. However, in the case of $\delta > 2$, we find, as expected, for some observables 
the sum over KK states can be divergent. We took an effective theory approach and cut the KK sum off 
at $M_*$, where we expect new physics to regulate this divergence. 

We take the view, as others have~\cite{Davoudiasl:2002fq}, that the standard three-active-flavor 
analysis explains
the neutrino oscillation data. (We do not address the LSND result.) This seems highly plausible 
given the Super-Kamiokande and the SNO neutral current data. Thus the oscillation into sterile KK
neutrino states must be constrained to be smaller than the current precision in experimental data.

Using the KK theory, we estimated the probability of the three active neutrino species to mix into 
sterile KK neutrino states. This probability is constrained by experiments such as CHOOZ and fits 
to the atmospheric oscillation data, leading to bounds on the radius ($R$) of extra dimensions that 
right-handed neutrinos propagate in. We have compiled the bounds on $1/R$ for $\delta = 1,2,3$ in 
Table.~\ref{OBND.TAB}, showing a strong bound for $\delta = 3$. For example, in the normal hierarchy of 
neutrino masses, we find $1/R$ has to be bigger than about $0.15,1.5,5600$~eV for $\delta=1,2,3$
respectively. We also present bounds for the inverted and the degenerate mass schemes. 
The mixing probability is divergent when summed over the KK states for $\delta\geq 2$, 
with a mild logarithmic dependence on the physics that comes in at $M_*$ for $\delta=2$, 
and a stronger linear dependence in the case of $\delta=3$. We do not predict the bounds for 
$\delta > 3$, since the dependence on the high energy completion of the extra dimensional theory we 
are working with would be increasingly stronger for bigger $\delta$. 
Our bounds have an additional source of uncertainty from using the continuum
approximation rather than performing a discrete sum over the KK states.

We also present bounds on $1/R$ coming from maintaining perturbative unitarity in the theory, 
in the Higgs-Higgs scattering channel. The bound is due to the growth in the large number of 
KK states that contribute in the intermediate state as the collision energy increases. 
The unitarity bound is also quite strong, especially for $\delta=3$, as shown in 
Fig.~\ref{fig:unitarity_bound}, albeit not as strong as the bound derived from oscillation data. 

It is interesting to ask what signatures of bulk neutrinos there might be at a high energy collider.
Since the only coupling to SM fields of the bulk neutrino is through the Yukawa coupling, 
one promising channel is the production of the KK states in the final state in association
with the Higgs. Even though the coupling of the Higgs to the neutrino is much suppressed,
the rate can be enhanced to measurable levels owing to the large number of KK states that 
can be produced in the final state. This is the subject of our forthcoming paper. 

\vspace*{5mm}
\noindent
{\bf Acknowledgments}~~~

We thank H.-J.~He, H.K.~Kim, T.~Tait, K.~Tobe and J.~Wells for useful discussions. This work was 
supported in part by the NSF grants PHY-0244919 and PHY-0100677.

\appendix
\section{2-component diagonalization}
\label{MDDS.APP}
We can arrive at the mass eigenvalues obtained in Sec.~\ref{MMD.SEC}, using two component Weyl fields   
and putting the neutrino mass matrix in a symmetric form. This is achieved by making the field 
definitions~\cite{Dienes:1998sb},
\beq
N^{(\hat n)} \equiv \frac{1}{\sqrt{2}} \left( \nu_R^{(\hat n)} + \nu_L^{(\hat n)} \right) , \ \ \ 
M^{(\hat n)} \equiv \frac{1}{\sqrt{2}} \left( \nu_R^{(\hat n)} - \nu_L^{(\hat n)} \right).
\eeq
We can then rewrite Eq.~(\ref{EQ4DL1.EQ}) as,
\bea
{\cal L}^{(4)} = -\frac{1}{2}\frac{|\hat n|}{R}\left[N^{(\hat n)} N^{(\hat n)} - M^{(\hat n)} M^{(\hat n)} + h.c.\right] - 
\frac{m_\nu}{v}h\left[\nu_R \nu_L + \sum_{\hat n}\left( N^{(\hat n)} \nu_L + M^{(\hat n)} \nu_L \right) +  h.c. \right].
\label{EQ4DL2.EQ}
\eea
Defining $\nu^T = (\nu_L\ \nu_R\ N^{(1)}\ M^{(1)}\ .\ .\ .\ N^{(N)}\ M^{(N)})$ and with 
$\left<h\right>=v$, we get the neutrino mass term,
\bea
{\cal L}^{(4)}_{\rm mass} = \frac{1}{2} \left[ \nu^T~{\cal M}~\nu + h.c. \right]
\eea
where the neutrino mass matrix ${\cal M}$ is given by,
\bea
{\cal M} = \pmatrix{
0&m_\nu&m_\nu&m_\nu&.&.&m_\nu&m_\nu&.&.&.&.\cr
m_\nu&0& & & & & & & & & & \cr
m_\nu& &\frac{1}{R}& & & & & & & & & \cr
m_\nu& & &-\frac{1}{R}& & & & & & & & \cr
.& & & &. & & & & & & & \cr
.& & & & &.& & & & & & \cr
m_\nu& & & & & &\left(\frac{|\hat n|}{R}\right)_{d_{\hat n}\times d_{\hat n}}& & & & & \cr
m_\nu& & & & & & &\left(-\frac{|\hat n|}{R}\right)_{d_{\hat n}\times d_{\hat n}}& & & & \cr
.& & & & & & & &.& & & \cr
.& & & & & & & & &.& & \cr
.& & & & & & & & & &\frac{N}{R}& \cr
.& & & & & & & & & & &-\frac{N}{R}
}.
\label{MMWY.EQ}
\eea
The alternating sign on the diagonal reflects the Dirac nature of masses. We have already noted 
that for $\delta > 1$, the state with mass $\pm\frac{|\hat n|}{R}$ at the ${\hat n}^{\rm th}$ level
has a degeneracy $d_{\hat n}$.

The characteristic equation, ${\rm Det}({\cal M}-\lambda~{\mathbf 1})=0$ is,
\bea
\left[ \prod_{|\hat n|=1}^{N} \left(|\hat n|^2 - (\lambda R)^2 \right)^{d_{\hat n}} \right] 
\left[(\lambda R)^2 - (m_\nu R)^2 + 
2 (\lambda R)^2 (m_\nu R)^2 \sum_{|\hat k|=1}^{N} \frac{d_{\hat k}}{|\hat k|^2-(\lambda R)^2} \right] = 0. 
\label{CHAREQN.EQ}
\eea
where, $k$ similar to $n$, is shorthand for $(k_1,...,k_\delta)$. In order to respect neutrino 
oscillation experimental bounds, we have to restrict ourselves to 
$m_\nu R \equiv \xi \ll 1$, and in the analysis that follows, we will keep terms only up 
to $O(\xi^2)$.

The lightest neutrino mass given by the lowest eigenvalue of Eq.~(\ref{CHAREQN.EQ}), in the 
limit $2\xi^2\sum_{\hat n}\frac{d_{\hat n}}{|\hat n|^2}\ll 1$, is 
\beq
m^{(0)} \approx \pm m_\nu (1-\xi^2\sum_{\hat n}\frac{d_{\hat n}}{|\hat n|^2}). 
\eeq

The heavier masses are as follows:\\
For $\delta=1$, we have $d_n = 1$ for all $n$ and this case was considered in Ref.~\cite{Dienes:1998sb}.
For the KK level with $d_n=1$, in Eq.~(\ref{CHAREQN.EQ}), $\lambda R=n$ is never a solution, 
and only the second of the two factors in square brackets is relevant. 
We find the eigenvalue $\lambda_n R = \pm n(1+\xi^2+\xi^4)$ and the corresponding eigenvector,
\beq
e_{n(1)}^\pm = \frac{1}{\sqrt{\cal N}_n}\pmatrix{1 & \frac{m}{\pm\lambda_n}& .& .& .& 
\frac{m}{(\pm\lambda_n-k/R)}& \frac{m}{(\pm\lambda_n+k/R)}& .& .& .}^T \ , 
\label{EN1.EQ}
\eeq
where the superscript $T$ on the row vector represents the transpose operation, and the reciprocal of the normalization constants are, 
\bea
\frac{1}{{\cal N}_0} = \frac{1}{2} \left[1-2\xi^2 \sum_{\hat n} \frac{d_{\hat n}}{\hat n^2} \right] \ , \ \ \ \ \ \frac{1}{{\cal N}_{\hat n}} = \xi_{\hat n}^2 d_{\hat n} . \nonumber
\eea
However, at the $n^{\rm th}$ KK level, if $d_n>1$ we can infer from Eq.~(\ref{CHAREQN.EQ}) that there are two classes of eigenstates:
\begin{itemize}
\item[$\bullet$] one with eigenvalue $\pm \frac{|n|}{R}\left[1+(\xi^2+\xi^4)d_n\right]$ with 
eigenvector as shown in Eq.~(\ref{EN1.EQ}), 
\item[$\bullet$] $(d_n - 1)$ states with eigenvalues $\pm \frac{|n|}{R}$ and the corresponding 
eigenvectors having the first $2n$ entries zero, and with the structure, 
\beq
e_{n(2,...,d_n)}^\pm = \pmatrix{0&.&.&0&v_1&.&.&v_{2d_n}&0&.&.&0}^T \ .
\eeq
\end{itemize}
The eigenvalues found here agree with Eqs.~(\ref{M0.EQ})~and~(\ref{MN.EQ}).

The matrix $V$ that diagonalizes the mass matrix Eq.~(\ref{MMWY.EQ}) have the $e_n$ as columns.
Neutrino oscillation from an active state to a sterile state computed using this matrix $V$ agrees
with that found using Eq.~(\ref{DL.EQ}). 

\section{Imaginary part of Passarino-Veltman functions}
\label{IPV.APP}
The unitarity of the $S-$matrix requires that 
twice of the imaginary part of any scattering amplitude can be written as 
\begin{equation}
2{\rm Im}\mathcal{M}(a\rightarrow b)=\sum_{f}\int d\Pi_{f}\mathcal{M}^{*}(b\rightarrow f)\mathcal{M}(a\rightarrow f), \label{eq:optical}\end{equation}
in which the sum $\sum_{f}$ runs over all possible 
final-state $f$, and $\Pi_{f}$ represents the phase space of final-state
$f$. We use the Cutkosky cutting rules to calculate the imaginary part
of the Passarino-Veltman scalar functions $B_0,C_0,D_0$ which were encountered in 
Sec.~\ref{UNICON.SEC}. 

\begin{figure}
\begin{center}
\scalebox{0.60}{\includegraphics{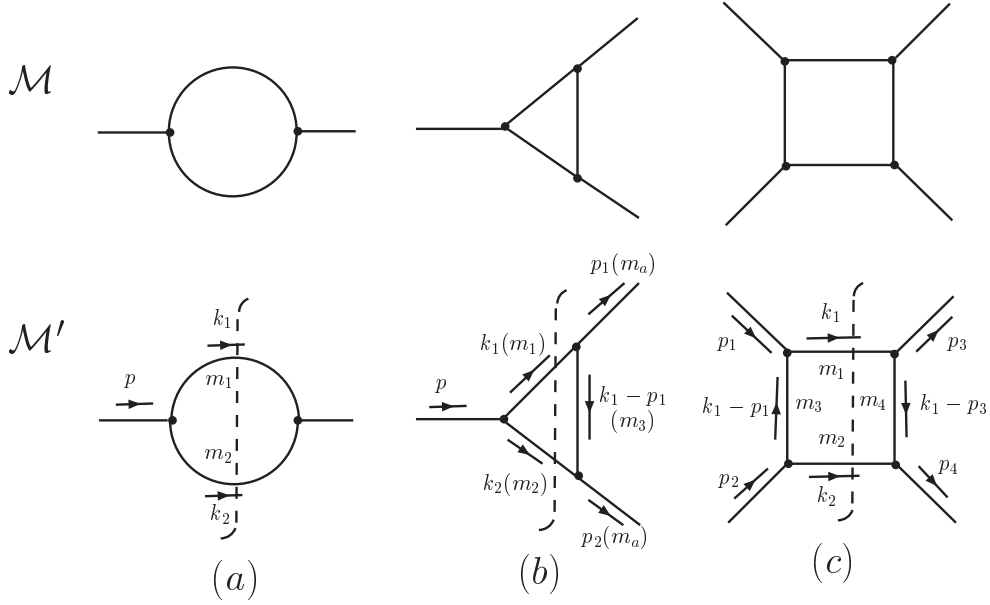}}
\caption{Feynman diagrams of scalar functions and the corresponding cut diagrams. \label{fig:img_bcd}}
\end{center}
\end{figure}

The Feynman diagrams ($\mathcal{M}_{i}$) and the
corresponding cut diagrams ($\mathcal{M}_{i}^{\prime}$), which are related
with each other as in Eq.~(\ref{eq:optical}), are shown in the upper
and lower parts of Fig.~\ref{fig:img_bcd}, respectively. The label $i=2,3,4$ represents
the two-point, three-point and four-point diagrams, respectively. In general
the N-point one-loop scalar integral in $D-$dimension is
\begin{equation}
T_{N}(p_{1},\cdots,p_{N-1},m_{0},\cdots,m_{N-1})=
\frac{(2\pi\mu)^{4-D}}{i\pi}
\int{\rm d}^{D}q\frac{1}{\mathcal{D}_{0}\mathcal{D}_{1}\cdots\mathcal{D}_{N-1}},
\label{eq:scalar}\end{equation}
with the denominator factors
\[
\mathcal{D}_{0}=q^{2}-m_{0}^{2},\qquad \mathcal{D}_{i}=(q+p_{i})^{2}-m_{i}^{2},\qquad
{\rm with}\;\; i=1,\cdots,N-1,
\]
originating from the propagators in the Feynman diagram. The $i\epsilon$
part of the denominator factors is suppressed. Here we adapt the conventions of Ref.~\cite{Denner:kt}.

\subsection{Two-point integrals, $B_{0}$ function}

The scalar two-point diagram ($\mathcal{M}_{2}$) in Fig.~\ref{fig:img_bcd}(a) is given by
\[
\mathcal{M}_{2}=\int\frac{{\rm d}^{4}q}{(2\pi)^{4}}\frac{1}{\mathcal{D}_{0}\mathcal{D}_{1}}=\frac{1}{16\pi^{2}}B_{0}(p^{2},m_{1}^{2},m_{2}^{2}),\]
from which we obtain the imaginary part of $B_{0}$, by applying the Cutkosky rules, as
\begin{eqnarray}
{\rm Im}(B_{0}) & = & 8\pi^{2}\times2\,{\rm Im(\mathcal{M}_{2})}\nonumber \\
 & = & 8\pi^{2}\times\mathcal{M}_{2}^{\prime}\nonumber \\
 & = & 2\,\int\frac{{\rm d}k_{1}}{2E_{1}}\frac{{\rm d}k_{2}}{2E_{2}}\delta^{4}(k_{1}+k_{2}-p)\nonumber \\
 & = & \pi\beta^{*},\label{eq:b0}\end{eqnarray}
where $\beta^{*}$ is defined as\[
\beta^{*}=\frac{1}{p^{2}}\sqrt{\left(p^{2}-(m_{1}+m_{2})^{2}\right)\left(p^{2}-(m_{1}-m_{2})^{2}\right)}.\]

\subsection{Three-point integrals, $C_{0}$ function}

\begin{figure}
\begin{center}
\scalebox{0.60}{\includegraphics{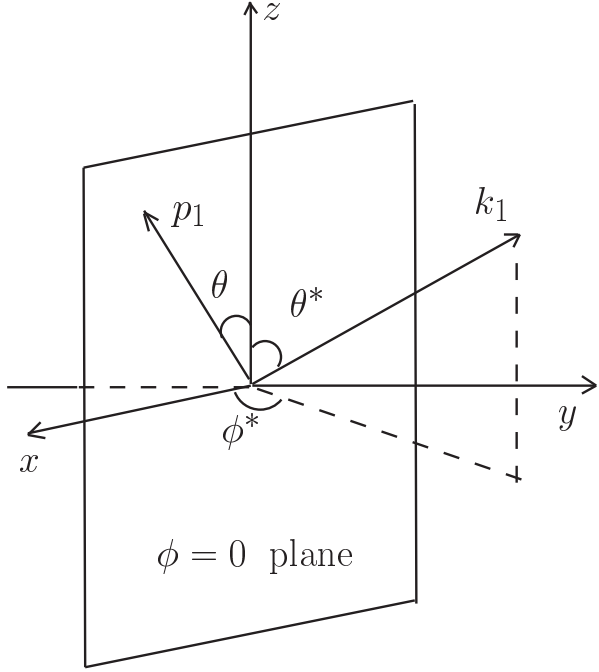}}
\caption{Kinematics of the cut diagrams \label{fig:kinematics_scalar}}
\end{center}
\end{figure}

The scalar three-point diagram ($\mathcal{M}_{3}$) in Fig.~\ref{fig:img_bcd}(b) is
given by\[
\mathcal{M}_{3}=\int\frac{{\rm d}^{4}q}{(2\pi)^{4}}\frac{1}{\mathcal{D}_{0}\mathcal{D}_{1}\mathcal{D}_{2}}=\frac{1}{16\pi^{2}}C_{0}(p_{1}^{2},p_{2}^{2},s,m_{1}^{2},m_{3}^{2},m_{2}^{2}),\]
from which we obtain the imaginary part of $C_{0}$ function as\begin{eqnarray*}
{\rm Im}(C_{0}) & = & 8\pi^{2}\times\mathcal{M}_{3}^{\prime}\\
 & = & \frac{1}{4}\beta^{*}\int{\rm d}\Omega^{*}\frac{1}{(k_{1}-p_{1})^{2}-m_{3}^{2}}.\end{eqnarray*}
In the c.m. frame, with the scattering angles $\theta,\phi,\theta^{*},\phi^{*}$ defined in 
Fig.~\ref{fig:kinematics_scalar}, the $4-$momenta is 
\begin{eqnarray}
p_{1} & = & \frac{\sqrt{s}}{2}(1,\beta\sin\theta,0,\beta\cos\theta),\nonumber \\
p_{2} & = & \frac{\sqrt{s}}{2}(1,-\beta\sin\theta,0,-\beta\cos\theta),\nonumber \\
k_{1} & = & \frac{\sqrt{s}}{2}(\hat{E_{1}},\beta^{*}\sin\theta^{*}\cos\phi^{*},\beta^{*}\sin\theta^{*}\sin\phi^{*},\beta^{*}\cos\theta^{*}),\nonumber \\
k_{2} & = & \frac{\sqrt{s}}{2}(\hat{E_{2}},-\beta^{*}\sin\theta^{*}\cos\phi^{*},-\beta^{*}\sin\theta^{*}\sin\phi^{*},-\beta^{*}\cos\theta^{*}),
\label{eq:kinematics}
\end{eqnarray}
with $s=p^{2}=(p_{1}+p_{2})^{2}$, $p_{1}^{2}=p_{2}^{2}=m_{a}^{2},$ $k_{1}^{2}=m_{1}^{2}$,
$k_{2}^{2}=m_{2}^{2}$ and \begin{eqnarray}
\beta & = & \sqrt{1-\frac{4m_{a}^{2}}{s}},\nonumber \\
\beta^{*} & = & \frac{\sqrt{\left( s-(m_{1}+m_{2})^{2}\right)\left( s-(m_{1}-m_{2})^{2}\right)}}{s},\nonumber \\
\hat{E_{1}} & = & \frac{s+m_{1}^{2}-m_{2}^{2}}{s},\nonumber \\
\hat{E_{2}} & = & \frac{s+m_{2}^{2}-m_{1}^{2}}{s}.\label{eq:kinematics2}\end{eqnarray}

After a straightforward calculation, we find the imaginary
part of $C_{0}$ to be  
\begin{equation}
{\rm Im}(C_{0})=\frac{\pi\beta^{*}}{p^{2}}\,\frac{1}{\beta\beta^{*}}\,\ln\frac{f+\beta\beta^{*}}{f-\beta\beta^{*}}\label{eq:c0}\end{equation}
where $f$ is defined as 
$$f=\frac{s-m_{1}^{2}-m_{2}^{2}-2m_{a}^{2}+2m_{3}^{2}}{s}.$$

\subsection{Four-point integrals, $D_{0}$ function}

The scalar four-point diagram ($\mathcal{M}_{4}$) in Fig.~\ref{fig:img_bcd}(c) is
given by\[
\mathcal{M}_{4}=\int\frac{{\rm d}^{4}q}{(2\pi)^{4}}\frac{1}{\mathcal{D}_{0}\mathcal{D}_{1}\mathcal{D}_{2}\mathcal{D}_{3}}=\frac{1}{16\pi^{2}}D_{0}(p_{1}^{2},p_{2}^{2},p_{3}^{2},p_{4}^{2},s,t,m_{1}^{2},m_{3}^{2},m_{2}^{2},m_{4}^{2}),\]
from which we obtain the imaginary part of $D_{0}$ function as\begin{eqnarray*}
{\rm Im}(D_{0}) & = & 8\pi^{2}\times\mathcal{M}_{4}^{\prime}\\
 & = & \frac{1}{4}\beta^{*}\int{\rm d}\Omega^{*}\frac{1}{(k_{1}-p_{1})^{2}-m_{3}^{4}}\,\frac{1}{(k_{1}-p_{3})^{2}-m_{4}^{2}}.\end{eqnarray*}
Using the same kinematics defined in Eqs.~(\ref{eq:kinematics}) and (\ref{eq:kinematics2}),
\bea
{\rm Im}(D_{0}) =
\left\{ 
 \begin{array}{l} 

     \displaystyle  \frac{2\pi\beta^{*}}{s^2}\;\frac{1}{\sqrt{K}}\;
	\ln \Biggl[ \frac{(BP-2AQ)(Px+Q)+2K-2P\sqrt{K(Ax^2+Bx+C)}}{Px+Q}\Biggl] \Biggl|^{x=1}_{x=-1}
              \\ [3mm]
     \hspace{11cm} ({\rm for}\; K > 0),
              \\ [7mm]
     \displaystyle  \frac{2\pi\beta^{*}}{s^2}\;\frac{1}{\sqrt{-K}}\;
	\arcsin \Biggl[ \frac{2K+(BP-2AQ)(Px+Q)}{P(Px+Q)\sqrt{B^2-4AC}}\Biggl] \Biggl|^{x=1}_{x=-1}\hspace{.8cm} ({\rm for} K < 0),
              \\ [7mm]
     \displaystyle -\frac{2\pi\beta^{*}}{s^2}\;
	\Biggl[ \frac{2P\;\sqrt{Ax^2+Bx+C}}{(BP-2AQ)(Px+Q)}\Biggl] \Biggl|^{x=1}_{x=-1}\hspace{4cm} ({\rm for}\; K = 0),

 \end{array}
\right. 
\eea
where the variables $A$, $B$, $C$, $P$, $Q$ and $K$ are defined to be
\bea
&&A=\beta^2 {\beta^{*}}^2, \\ \nonumber
&&B=-2f\beta\beta^{*}\cos \theta, \\ \nonumber
&&C=f^2-\beta^2{\beta^{*}}^2\sin^{2}\theta, \\ \nonumber
&&P=\beta \beta^{*}, \\ \nonumber
&&Q=-f, \\ \nonumber
&&K=AQ^{2}-BPQ+CP^2,
\eea
and $f$ is defined as
$$f=\frac{s-m_{1}^{2}-m_{2}^{2}-2m_{a}^{2}+2m_{3}^{2}}{s}.$$


\end{document}